\documentclass[twocolumn,10pt]{article}
 \usepackage{usenix2019_v3}

\usepackage{verbatim}
\usepackage{amsmath}
\usepackage{amssymb}
\usepackage{algorithmic}
\usepackage{algorithm}

\usepackage{graphicx}
\usepackage{epstopdf}
\usepackage{array}
\usepackage{booktabs}
\usepackage{multirow}
\usepackage{caption} 
\usepackage{multicol}
\usepackage{tabularx}
\usepackage{framed}
\usepackage{enumitem}
\usepackage{threeparttable}
\usepackage[compact]{titlesec}

\graphicspath{{figs/}}

\usepackage{xcolor}
\usepackage{colortbl} 
\definecolor{jade}{rgb}{0.0, 0.66, 0.42}
\definecolor{carolinablue}{rgb}{0.6, 0.73, 0.89}
\definecolor{dkgreen}{rgb}{0,0.6,0}
\definecolor{dkblue}{rgb}{0,0.4,0.5}
\definecolor{gray}{rgb}{0.5,0.5,0.5}
\definecolor{mauve}{rgb}{0.58,0,0.82}

\definecolor{codegreen}{rgb}{0,0.6,0}
\definecolor{codegray}{rgb}{0.5,0.5,0.5}
\definecolor{codepurple}{rgb}{0.58,0,0.82}
\definecolor{codeblue}{rgb}{0,0,205}
\definecolor{backcolour}{rgb}{245,245,245}

\usepackage{listings}
\usepackage{enumitem}

\lstdefinelanguage
   [x64]{Assembler}     
   [x86masm]{Assembler} 
   {morekeywords={xend, CDQE,CQO,CMPSQ,CMPXCHG16B,JRCXZ,LODSQ,MOVSXD, %
                  POPFQ,PUSHFQ,SCASQ,STOSQ,IRETQ,RDTSCP,SWAPGS, %
                  rax,rdx,rcx,rbx,rsi,rdi,rsp,rbp, %
                  r8,r8d,r8w,r8b,r9,r9d,r9w,r9b, %
                  r10,r10d,r10w,r10b,r11,r11d,r11w,r11b, %
                  r12,r12d,r12w,r12b,r13,r13d,r13w,r13b, %
                  r14,r14d,r14w,r14b,r15,r15d,r15w,r15b}} 

\lstdefinestyle{mystyle}{
    backgroundcolor=\color{backcolour},   
    commentstyle=\color{codegreen},
    keywordstyle=\color{codeblue},
    numberstyle=\tiny\color{codegray},
    stringstyle=\color{codeblue},
    basicstyle=\ttfamily\footnotesize,
    breakatwhitespace=false,         
    breaklines=true,                 
    captionpos=b,                    
    keepspaces=true,                 
    numbers=left,                    
    numbersep=5pt,                  
    showspaces=false,                
    showstringspaces=false,
    showtabs=false,                  
    tabsize=2
}
\usepackage{url}

\usepackage{flushend}
\usepackage{balance}

\usepackage[misc]{ifsym}
\usepackage{bbding}

\usepackage{xspace}

\usepackage{xspace}
\newcommand{\sysname}{PACSan\xspace}

  \newcommand{\runperf}{\textcolor{black}{ 7.172\%}\xspace}
  \newcommand{\memperf}{\textcolor{black}{ 89.063\%}\xspace}

  \newcommand{\runoverhead}{\textcolor{black}{ 0.84$\times$}\xspace}
  \newcommand{\memoverhead}{\textcolor{black}{ 1.92$\times$}\xspace}

 \usepackage[markup=underlined]{changes}

\begin{document}

\date{}

\title{\sysname: Enforcing Memory Safety Based on ARM PA}


\author{
{\rm Yuan Li} \\
{\rm Tsinghua University}
\and 
{\rm Wende Tan} \\
{\rm Tsinghua University}
\and
{\rm Zhizheng Lv} \\
{\rm Tsinghua University}
\and
{\rm Songtao Yang} \\
{\rm Tsinghua University}
\and
{\rm Mathias Payer} \\
{\rm EPFL}
\and
{\rm Ying Liu} \\
{\rm Tsinghua University}
\and
{\rm Chao Zhang} \\
{\rm Tsinghua University}
}

\maketitle

\thispagestyle{empty}

\begin{abstract}
Memory safety is a key security property that stops memory corruption vulnerabilities.
Existing sanitizers enforce checks and catch such bugs during development and testing.
However, they either provide partial memory safety or have overwhelmingly high performance overheads.

Our novel sanitizer \sysname enforces spatial and temporal memory safety with no false positives at low performance overheads.
\sysname removes the majority of the overheads involved in pointer tracking by
sealing metadata in pointers through ARM PA (Pointer Authentication),
and performing the memory safety checks when pointers are dereferenced.

We have developed a prototype of \sysname and systematically evaluated its security and performance on the {\tt Magma}, {\tt Juliet}, {\tt Nginx}, and {\tt SPEC CPU2017} test suites, respectively.
In our evaluation, \sysname shows no false positives together with negligible false negatives, while introducing stronger security guarantees and lower performance overheads than state-of-the-art sanitizers,
including HWASan, ASan, SoftBound+CETS, Memcheck, LowFat, and PTAuth.
Specifically, \sysname has \runoverhead runtime overhead and \memoverhead memory overhead on average.
Compared to the widely deployed ASan, \sysname has no false positives and much fewer false negatives, and reduces \runperf runtime overheads and \memperf memory overheads.
\end{abstract}


    


\section{Introduction}

Memory safety violations are still the most common root cause of modern exploits~\cite{Google,Miller}.
In practice, there are two types of memory safety violations~\cite{sok-sanitizer}:
(1) spatial safety violations where a program performs an out-of-bound memory access, e.g., buffer overflow and buffer under-read,
and (2) temporal safety violations where a program accesses memory of invalid state (i.e., unallocated or freed), e.g., use-after-free and double free.
Such memory safety violations break the integrity of memory and program states, which can further lead to denial of service, sensitive data leakage and corruption, privilege escalation, or even control-flow hijacking.

Many sanitizers have been proposed to enforce memory safety and catch memory safety violation bugs.
For instance, AddressSanitizer (ASan)~\cite{asan} detects buffer overflow and use-after-free bugs, by quarantining freed objects and padding active objects with non-accessible redzones and tracking the accessibility of each memory byte.
ASan is the most widely deployed sanitizer used in fuzz testing, e.g., by AFL~\cite{afl}.
However, these sanitizers in general are limited.

On the one hand, most sanitizers only provide partial memory safety guarantees. 
A full memory safety solution should guarantee that,
each memory object can only be accessed within its bounds (spatial safety) and must be in a valid state when accessed (temporal safety).
But existing solutions fail to enforce this property.
For instance, SoftBound~\cite{softbound} and LowFat~\cite{lowfat-heap,lowfat-stack} only provide spatial memory safety.
CETS~\cite{cets} and the recently proposed ARM PA-based solution PTAuth~\cite{ptauth} only provide temporal memory safety, while PTAuth exclusively protects heap objects (ignoring the stack and global variables).
Even for the most widely deployed sanitizer---ASan, it also has many false negatives, i.e., it fails to detect certain memory safety violations that miss its redzones
as demonstrated by MEDS~\cite{meds}. 

On the other hand, sanitizers that aim at providing full memory safety guarantees, in general, have overwhelmingly high performance overheads.
ASan has 0.904$\times$ runtime overhead and 17.55$\times$ memory overhead on average.
HWASan~\cite{HWASAN} utilizes hardware features to lower the memory overhead, but still suffers from 1.33$\times$ runtime overhead.
Memcheck~\cite{nethercote2007valgrind} yields 20.36$\times$ runtime overhead and 6.11$\times$ memory overhead on average.
The combination of SoftBound~\cite{softbound} and MPX~\cite{mpx} yields 7.09$\times$ memory overhead.
The overwhelmingly high performance overhead renders sanitizers inefficient, greatly slowing down the process of debugging, fuzzing, and other applications.

To enforce full memory safety, as discussed by previous studies~\cite{sok-sanitizer,asan}, we must track properties (metadata) of each object, including 
(1) its base address and size (spatial property) and (2) its birthmark (temporal property),
and perform property checks when objects are accessed via pointers.
The key to the success of such solutions is how to \emph{efficiently} track properties (or metadata) and perform property checks {\em without lowering} security guarantees.
In general, overheads of sanitizers come from four types of operations: creating metadata (at the time of object creation, i.e., \textit{alloc}), propagating metadata (at the time of pointer operation, i.e., \textit{ptr\_x=ptr\_y}), checking metadata (at the time of object access, i.e., \textit{*ptr\_x}), and cleaning up metadata (at the time of object deallocation, i.e., \textit{free}).
The metadata propagation operation is the most time-consuming one (as shown in LowFat~\cite{lowfat-stack}), 
which could be optimized to improve the performance. For instance, ASan~\cite{asan} only tracks metadata per object instead of per pointer, which shows better performance than its precedent solutions.
But its overhead is still high, and it cannot provide full memory safety guarantees.

In this paper, we propose a novel sanitizer \sysname to enforce  spatial and temporal memory
safety at low overheads with no false positives and negligible false
negatives.
%
\sysname improves performance by eliminating metadata propagation in a clever way, by encoding the metadata into pointers using
the hardware feature ARM PA (Pointer Authentication~\cite{pa}).
Specifically, we first create and place objects' metadata in a linear table, and
utilize ARM PA to generate a PAC (Pointer Authentication Code) signature for each object.
The PAC is stored in the high-order bits of pointers associated with the object, which is implicitly propagated no matter how the pointer is used (e.g., pointer assignment, pointer arithmetic operation, and function argument passing), thus saving the overhead of metadata propagating.
Furthermore, the PAC signature in each pointer serves as an index to the metadata table, which enables efficient metadata lookup and greatly reduces the overhead of metadata checking, together with the signature verification support provided by ARM PA.

On the other hand, \sysname has no false positives, since it precisely tracks the full memory safety properties and performs precise spatial and temporal safety checks.
In addition, \sysname has negligible false negatives, much lower than existing solutions.
The only source of false negatives comes from sub-object overflow, since no metadata is tracked for fields of objects due to performance reason (as most sanitizers did).
Specifically, it provides stronger defenses than ARM PA and its succedent ARM MTE (memory tagging extension).

We have implemented a prototype of \sysname on the ARM64 architecture in a real device.
Then, we systematically evaluated its security on the {\tt Magma}~\cite{Magma} and {\tt Juliet}~\cite{Juliet} benchmark, and evaluated its performance on {\tt SPECspeed 2017} and {\tt Nginx}.
Evaluation results show that, \sysname provides stronger security guarantees than state-of-the-art sanitizers,
including HWASan, ASan, SoftBound+CETS, Memcheck, LowFat, and PTAuth, while introducing lower performance overheads.
Specifically, \sysname has \runoverhead runtime overhead and \memoverhead memory overhead on average.
Compared with ASan, \sysname has much lower false negatives and exhibits \runperf lower runtime overheads and \memperf lower memory overheads.
Furthermore, it shows that ARM PA instructions have a big performance gap between the ARM FVP emulator and real devices.

In summary, we make the following contributions:
\begin{enumerate}[noitemsep,nolistsep]

\item We propose a novel sanitizer \sysname which utilizes the ARM PA feature to enforce full memory safety stronger than ARM PA and ARM MTE.

\item
We propose to use PAC signatures to eliminate the overhead of metadata propagation and enable efficient runtime metadata lookup and checks, therefore greatly reduces the runtime performance overheads.

\item We implement a prototype in a real hardware device, and show a huge performance gap between the ARM FVP emulator and real devices for the first time. 

\item We evaluate \sysname systematically in terms of security, runtime and memory overhead, and show that it outperforms state-of-the-art sanitizers. 
\end{enumerate}

\section{Background}

\subsection{Memory Safety}

 Memory safety violations fall into two categories~\cite{sok-memory}: 
 (1) {\em spatial violation}s happen when a pointer accesses out of its referent object's bound, e.g., buffer overflow, and
 (2) {\em temporal violation}s happen when a pointer accesses an invalid object (unallocated or freed), e.g., double free and use after free.
 In addition to memory safety bugs, programs may have several other types of bugs, including uninitialized variables, type cast bugs, misuse of functions with variable arguments (e.g., format string), or other types (e.g., integer overflow, command injection etc.), which are out of the scope of this paper.

 {\bf Spatial Memory Safety.}
 Spatial corruption refers to out-of-bounds accesses.
 Buffer overflow is the most common type of spatial corruption vulnerabilities.
 It has a long story of being studied by adversaries and exploited to launch attacks~\cite{rop}.
 Access bound checking is the most effective solution to detect (and prohibit) spatial corruptions.
 ASan~\cite{asan} uses redzones around objects to stop out-of-bound violations by checking if memory accesses target redzones.
 HWASan~\cite{HWASAN} tags each allocated memory block and its pointer,
 and utilizes the AArch64 hardware feature to store the pointer tag in the pointer. If the tag of the pointer mismatches the refered object's, an access violation is caught.
LowFat~\cite{lowfat-heap,lowfat-stack} also encodes the bound information into the pointer, with a special encoding scheme. 
 So LowFat can utilize the pointer to retrieve the bound information before each pointer dereference, to check if that memory access operation is within the bound.
 
 {\bf Temporal Memory Safety.}
 Generally, temporal memory safety violations are caused by programs accessing unallocated or deallocated objects.
 When programs explicitly deallocate an object, the object becomes invalid, and all pointers to this object become \textit{dangling pointers}.
 When a dangling pointer is freed or used again, a double-free or use-after-free (UAF) bug is yielded. 
 Tag matching is a common method used by mitigations~\cite{patil1997low,austin1994efficient,cets} to catch temporal safety violations.
 In general, they assign tags to objects and pointers, and compare them at pointer dereferences.
 Another common method is detecting dangling pointers before they are used. 
 Specifically, such methods will monitor and track pointers passed to the \texttt{free} function.
 If a marked pointer is used later, a temporal memory violation is reported.
 However, this method cannot handle copies of dangling pointers.
 Thus, some tools not only mark pointers to be freed but also maintain an object-to-pointer map to invalidate copies of dangling pointers when \texttt{free} is called, such as Undangle~\cite{Undangle}, DangNull~\cite{DangNull}, FreeSentry~\cite{FreeSentry}, and DangSan~\cite{dangsan2017}.

\begin{figure}[t]
  \centering
    \includegraphics[width=0.33\textwidth]{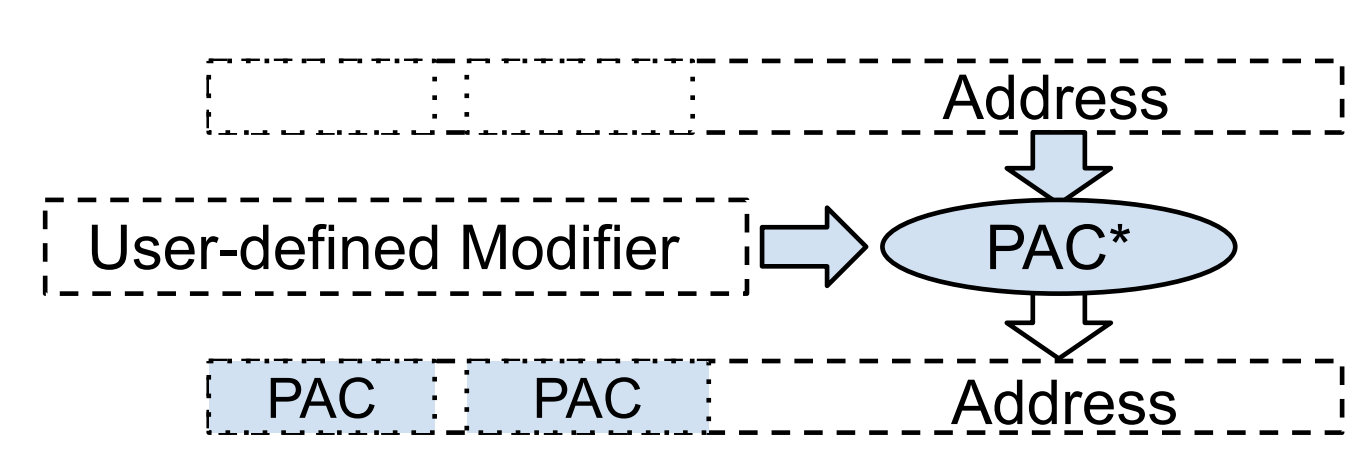}
    \caption{The PAC is generated by \texttt{PAC*}-family instructions, using the target pointer, a modifier, and a key (in kernel).} 
    \label{fig-PAC-compute}
\end{figure}

\subsection{ARM Pointer Authentication}

Armv8.3-A introduces the PA (Pointer Authentication)~\cite{pa} security extension, which has been applied in recent iOS devices~\cite{ios}.
This extension enforces pointer integrity by signing the pointer at each definition point and verifying the signature when the pointer is used.

As shown in Figure \ref{fig-PAC-compute},  \texttt{PAC*} instructions sign a target pointer and compute its Pointer Authentication Code (PAC), which is stored in unused high-order bits of the target pointer.
Further, \texttt{AUT*} instructions authenticate a signed pointer.
If the authentication succeeds, the PAC embedded in the pointer is stripped, and the pointer can be dereferenced as normal.
Otherwise, the pointer will be modified to an invalid pointer in Armv8.3-A or will trigger an exception in ARMv8.6-A.
We assume PA has the latter behavior in this paper.

Users of ARM PA could choose modifiers as wish to tune signatures, so that a same pointer can yield different signatures in different execution contexts.
The modifier also serves as a bound between the pointer definition point and verification point, since it has to be the same in order to pass the check.

\begin{figure}[t]
    \centering
    \includegraphics[width=0.45\textwidth]{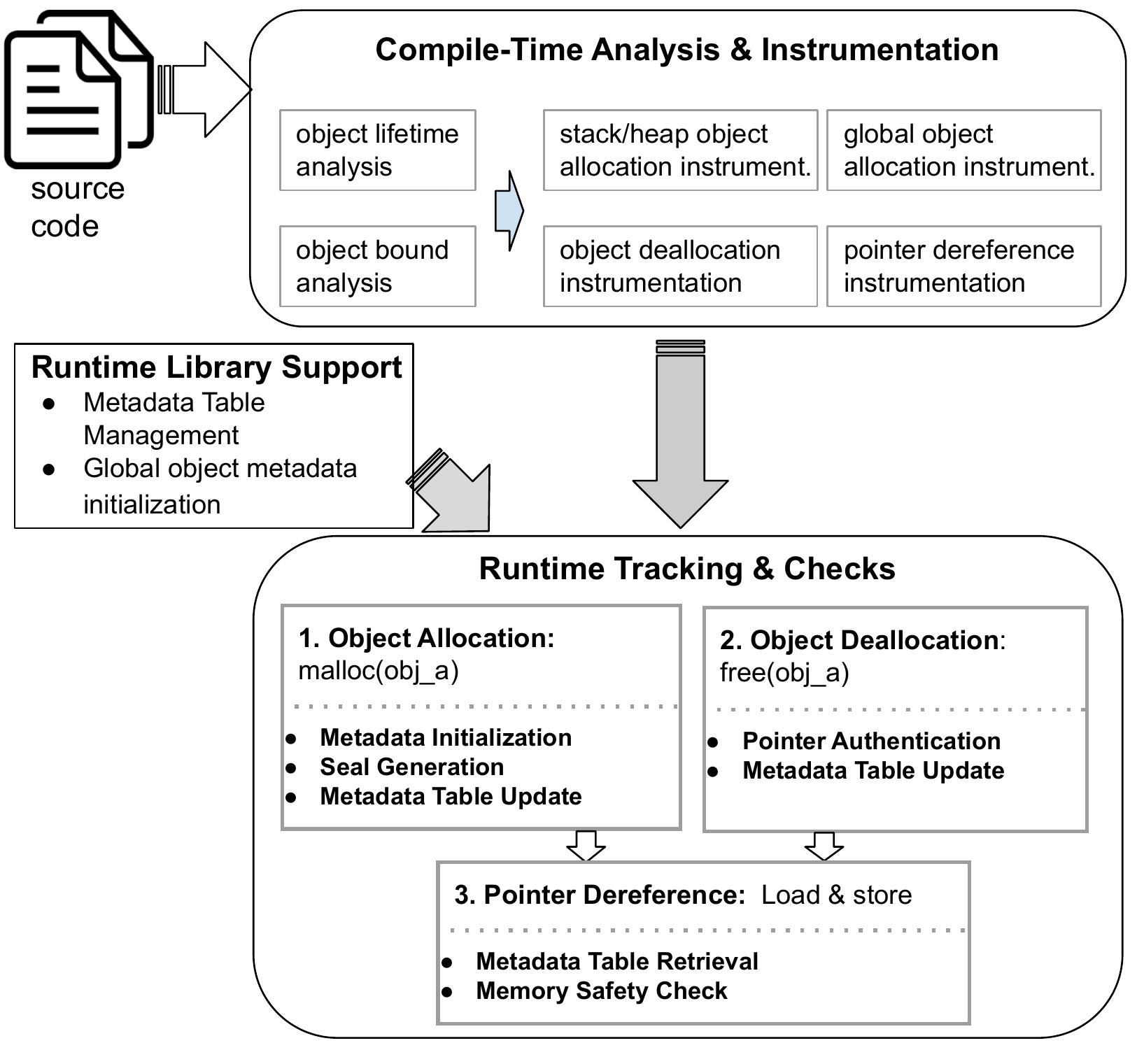}
    \caption{Overview of \sysname. It updates metadata at object allocation and deallocation sites, seals the metadata into pointers to avoid propagation overhead, and checks metadata at each pointer dereference  using the seal to retrieve metadata.
    } 
    \label{fig-PACMem-overview}
\end{figure}

\section{Methodology}

\subsection{System Overview}
\sysname is designed for efficiently catching memory safety bugs in target programs at runtime, with negligible false negatives. 
Following common practice, it also tracks metadata of objects and pointers, and performs metadata  checks  before memory accesses. 
To enforce full memory safety, \sysname tracks {\em all} necessary spatial and temporal metadata precisely, and checks {\em all} memory accesses.

In general, metadata is created when objects are allocated, is propagated to other pointers during pointer operations, is checked when pointers are used to access objects, and is cleaned up when objects are deallocated.
Each of these four steps would cause runtime overheads, among which the metadata propagation and checking are the most time consuming.
\sysname utilizes ARM PA to eliminate metadata propagation and accelerate metadata checking.

Figure~\ref{fig-PACMem-overview} demonstrates the overview of \sysname.
At compile time, \sysname analyzes the bound and liveness properties of objects (including global objects), and instruments at proper locations to  track such metadata or perform memory safety checks. Specifically, \sysname generates metadata when an object is created, utilizes ARM PA to generate a PAC signature (denoted as a {\em seal}) of this metadata, which will be embedded into pointers associated to this object,
and places the metadata in a table indexed by this seal.
The seal will implicitly propagate along with pointers no matter how they are used, saving overheads of the metadata propagation.
Furthermore, 
before a pointer is dereferenced, the seal serves as an index to efficiently lookup the metadata table, 
enabling efficient metadata checks with the support of signature verification provided by ARM PA.
Lastly, when an object is deallocated, the object's pointer will be checked as in regular pointer dereference operations, and the metadata will be removed from the metadata table once the check succeeds.

\subsection{Metadata Creation}
\label{sec:design:metadata}
To enforce full memory safety, we track all necessary memory safety metadata.
In \sysname, each object's metadata consists of (1) its base address, (2) its object size, and (3) its birthmark,
while the former two are used for spatial memory safety checks and the last one for temporal memory safety checks.
Specifically, the metadata takes 128 bits, whereas the birthmark and object size take 32 bits respectively.

\begin{figure}[t]
     \centering
    \setlength{\abovecaptionskip}{2mm}
    \includegraphics[width=.45\textwidth]{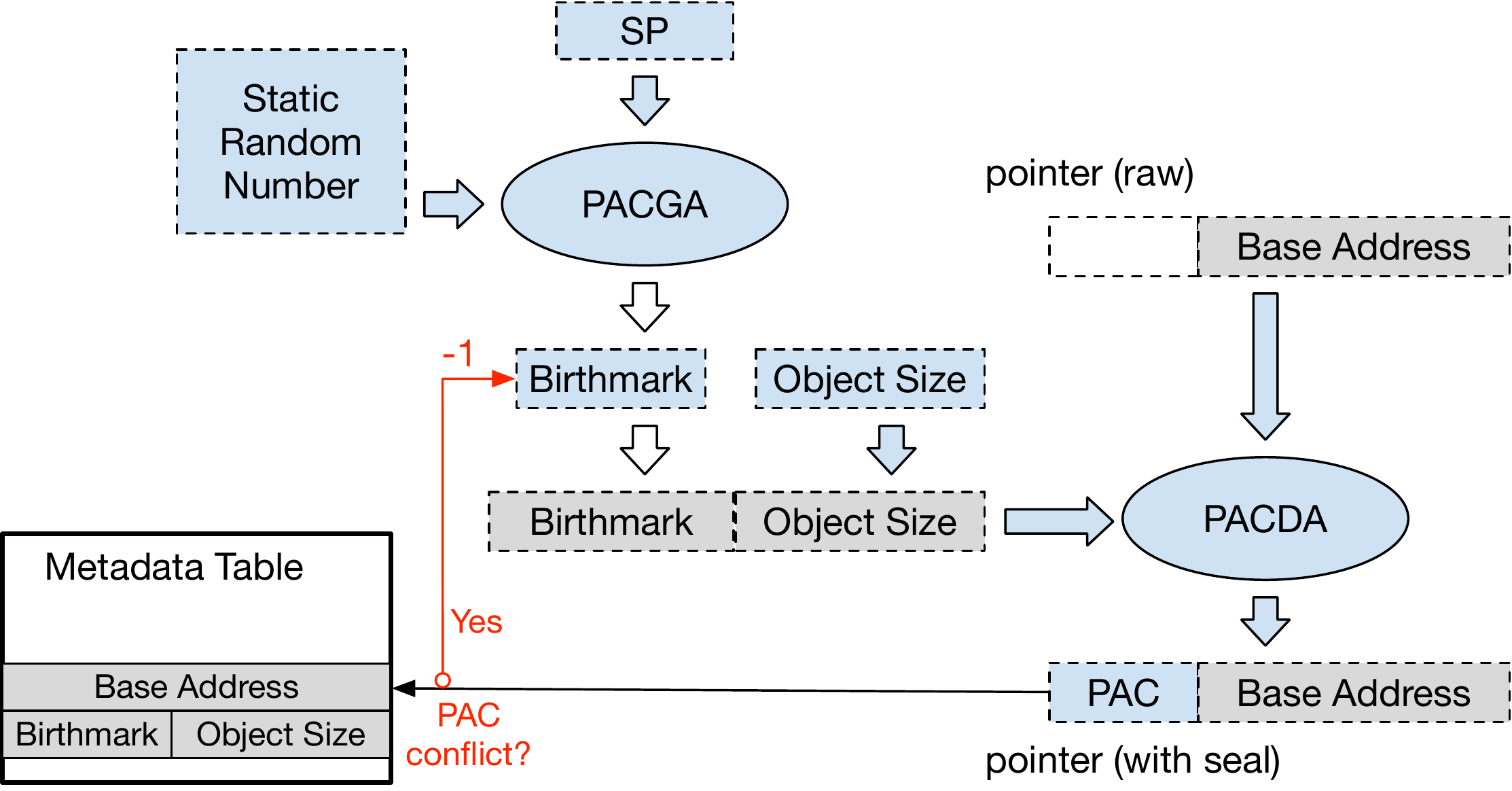}
    \caption{Metadata creation: when a memory object is allocated, a random birthmark is yielded to compose its metadata, which will be further signed to generate the pointer's seal and be placed in the metadata table using the seal as the index.
    } 
    \label{fig-PACMem-allocate}
\end{figure}

As shown in Figure~\ref{fig-PACMem-allocate}, the metadata is stored in a metadata table, i.e., a linear array for performance consideration.
To retrieve object metadata, it is common for sanitizers to directly use the associated pointer as the index, as ASan does.
However, such solutions in general will yield very high memory overheads, since the table size is proportional to the pointers' value space.
We propose to seal metadata into the high-order bits of pointers (via the hardware feature ARM PA), and use the seal as the index to retrieve objects' metadata,
which greatly reduces the memory size of the metadata table and also enables efficient runtime metadata retrieval.

As shown in Figure~\ref{fig-PACMem-allocate}, the seal is a PAC signature of the metadata, and  thus has a special bond with the pointer's metadata, which can be verified at runtime.
Specifically, for each newly allocated object, we assign it with a pseudo random birthmark.
But as ARM has no architecture support for random number generation, we again utilize ARM PA to generate a pseudo random birthmark, by taking the current dynamic stack pointer  as the pointer and taking a static random number generated at compile time as the modifier.
Then, we utilize ARM PA to yield the seal for the metadata (i.e., base address and size, and the birthmark).\footnote{The seal and metadata initialization should be done when objects are allocated. For global objects without explicit allocation sites, we utilize initializer functions to generate seal and metadata for them at program startup. }
At runtime, metadata indexed by this seal should be consistent with it.

Ideally, two different objects should have different metadata and seals.
However, due to the limitation of digital signature, two different metadata could yield a same seal. As a result, two metadata will be stored at the same slot in the metadata table, and at least one runtime metadata check will fail and cause false positives when these two objects are both accessed.
To ensure the uniqueness of each object's seal, 
the metadata initialization code instrumented by \sysname will repeatedly change the modifier to yield different seals until a non-conflicting one is found.
Specifically,  if the seal of a newly created object collides with an existing object's (i.e., the slot in the metadata table is taken), the modifier (i.e., birthmark) decreases by  one and a new seal will be yielded. 
Note that, this collision mitigation overhead only exists when a new object with a conflicted seal is created. 
After creating the object, all seals are conflict-free and can be used  (e.g., to retrieve metadata from the table) without extra overheads.

\begin{figure}[t]
  \centering 
     \includegraphics[width=.42\textwidth]{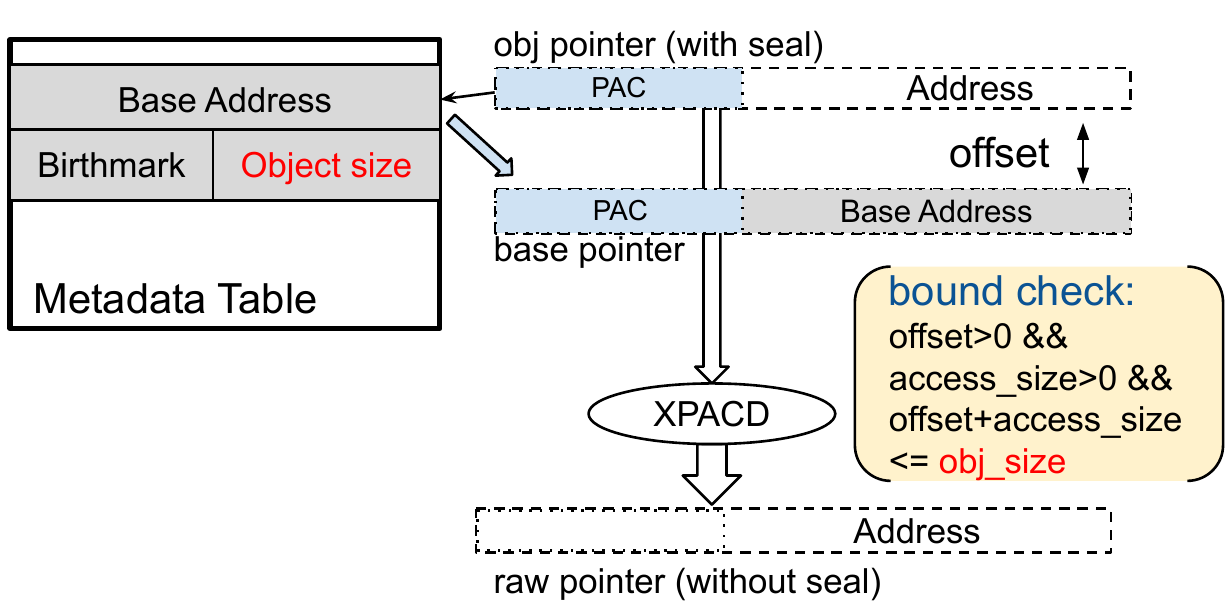}
     \caption{Metadata checking: when a pointer is dereferenced, \sysname retrieves metadata from the metadata table, calculates the pointer's offset to its base address, and performs a delicate bound checks to enforce full memory safety.} 
     \label{fig-PACMem-check}
 \end{figure}

Compared to existing works, \sysname's metadata management scheme has several advantages.
First, it tracks all temporal and spatial properties, and enables full memory safety enforcement.
Second, it avoids the metadata propagation overheads. 
The indices of the metadata table (i.e., seals) are shipped together with pointers, and are not affected by pointer operations (e.g., pointer arithmetic operations).
Thus, we do not need to perform extra metadata propagation like Watchdog~\cite{watchdog} did.
Third, the metadata retrieval is simple and lightweight, i.e.,
the seal can be used as an index to retrieve data from a linear array (i.e., the metadata table).
In addition, \sysname has significantly lower memory overheads than other methods (e.g., ASan and HWASan) due to the compact design of metadata and seals.

\subsection{Metadata Tracking and Checking}\label{sec:workflow}
Once the metadata (or seal) is created for an object (or pointer), the seal will propagate along with the pointer automatically.
When the pointer is dereferenced (i.e., memory access or object deallocation), its metadata will be checked.

\subsubsection{\bf Memory Access Checking}

Whenever a memory address is loaded through a pointer, a security check must enforce full memory safety.
As shown in Figure~\ref{fig-PACMem-check},
\sysname adopts a bound check to enforce spatial memory safety.
Specifically, it uses the pointer's seal as the index to retrieve  base address and size of the intended object from the metadata table.
With this base address, we can calculate the offset of the pointer being dereferenced, and verify that the memory access is in valid bounds.

Note that this bound check also provides temporal safety guarantees.
If the pointer being dereferenced is a dangling pointer, the aforementioned bound check will (likely) fail.
Since the dangling pointer's intended object has been freed, its metadata entry in the metadata table is cleared.
In the first case, the cleared entry has not been taken by other objects yet,
then the retrieved base address and object size are all 0, and the bound check will fail.
Furthermore, the probability of a hash collision for use-after-free is approximately $5.96 \times 10^{-8}$, i.e., vilations are detected with over 99.9999999\% probability.
Even though the corresponding entry is taken by another object which accidentally has the same seal,
the retrieved base address and object size are unlikely the same as the freed object's, and the bound check will fail too.
As a result, this bound check is sufficient to catch temporal safety bugs, including use-after-free (UAF) bugs.


\begin{figure}[t]
     \centering
    \includegraphics[width=0.4\textwidth]{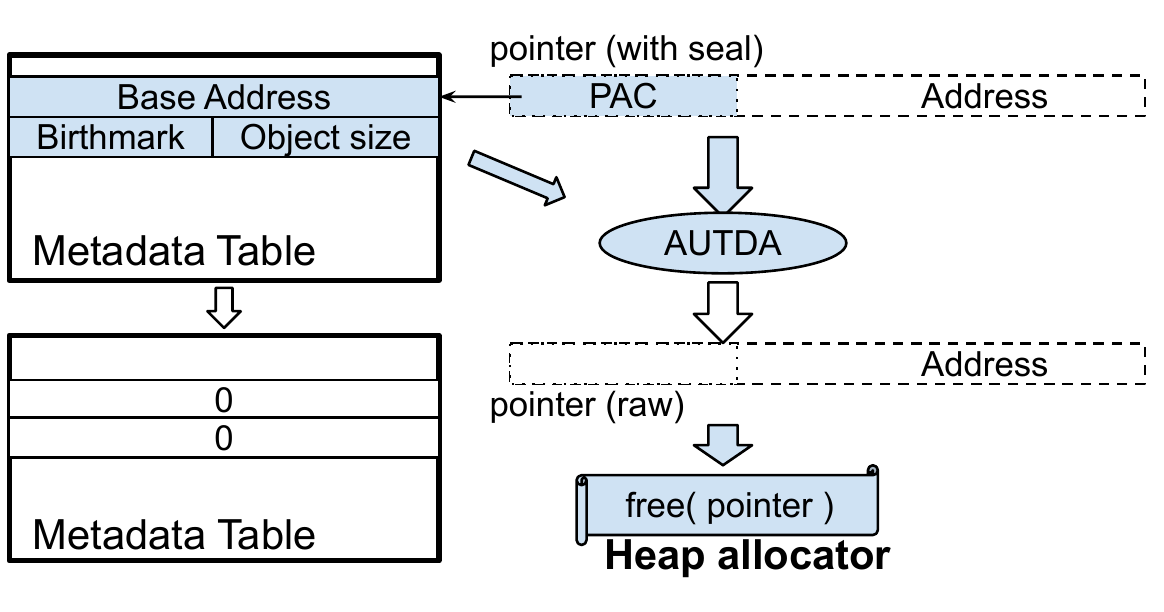}
    \caption{Object deallocation phase: clear the metadata table, and authenticate the pointer if it is a heap object.}
    \label{fig-PACMem-free}
\end{figure}

\subsubsection{\bf Object Deallocation Checking}
When a heap or stack object is freed (explicitly deallocated or implicitly purged), we need to remove its outdated metadata from the metadata table, i.e., set the table entry to zeros.

If the object to be freed is a heap object, we make an extra check before clearing the metadata table.
Specifically, \sysname retrieves the birthmark and object size from the metadata, and uses it to authenticate the seal of the pointer using ARM PA instruction \texttt{AUTDA}, and sends the raw pointer to the heap allocator to free.
In this way, it stops the heap allocator from freeing invalid pointers, and catches double-free and invalid-free bugs.
Figure~\ref{fig-PACMem-free} outlines this workflow.

\subsection{Compatibility with unprotected modules}
Since \sysname utilizes upper bits of pointers to store seals, it would cause compatibility issues if pointers are used across protected and unprotected modules. 
Therefore, \sysname takes extra steps to deal with such compatibility issues.

First, when a signed pointer is passed to a unprotected module, \sysname will do a safety check for it (i.e., verify the pointer and remove PAC).
This will eliminate compatibility issues, and by the way stops potential dangling pointers from being used in unprotected modules.

Second, when an non-signed pointer is yielded by unportected modules and returned to protected modules,\footnote{If there are recursive pointers in the objects, \sysname could follow the API definition to get all those pointers and sign them recursively.} \sysname will generate a conservative metadata (i.e., minimum base address and maximum object size) for it, since it does not know how the object is created.
This addresses the compatibility issue but leaves an attack surface for adversaries. 
However, this is inevitable for unprotected modules.
To provide full protection, developers are strongly encouraged to protect all modules with \sysname.

%



\section{System Implementation}

\subsection{Implementation details}
The \sysname system prototype includes (i) a custom compiler extension for analysis and instrumentation and (ii) a runtime support library for creating the metadata table and initializing metadata for global objects.
The current prototype supports C and C++ programs. 
It supports all common memory allocation APIs such as \texttt{malloc},
\texttt{free}, \texttt{new}, and \texttt{delete}.

Our compiler extension is based on LLVM 14.0.0, which already supports ARM PA instructions. 
We first analyze the target program using a new LLVM pass, 
and insert the security checks mentioned in Section~\ref{sec:workflow} at proper locations.
It has to work with the \texttt{-fPIC} compiler option to make target programs position-independent, so that global objects are all referenced via the \texttt{GOT} (global offset table) and easy to recognize.

Our runtime library mainly allocates a metadata table in shadow memory before the protected program starts.
We use a constructor function which will run at program startup to initialize the metadata of all global variables and put seals of the pointers in the \texttt{GOT}.
Then, we make the \texttt{GOT} read-only to prevent the \texttt{GOT} from being corrupted by attackers. 

Each capability corresponds to a total of 16 bytes of metadata, including 8-bytes signed base address and 8-bytes modifier (including 4-bytes birthmark and 4-bytes  object size).

\subsection{Performance Optimizations}
To lower the overheads of \sysname, we have made the following optimizations without lowering security guarantees.


{\bf Loop-Independent Memory Operations. }
If a memory operation in a loop accesses a same memory range no matter how many times the loop has iterated,
\sysname just checks once for it before the loop entry.

{\bf Loop Bound Pointers.  }
If a pointer only increases or decreases linearly within a loop, and we can statically determine its upper bound and lower bound at compile time,
then \sysname will skip checks for dereferences to the pointer within this loop and safely performs checks for the upper bound and lower bound before the loop entry.

{\bf Redundant Check Elimination.}
If one memory access instruction dominates or post-dominates another one, and their accessed address ranges are the same, the checks for the second instruction are considered redundant and could be removed. 
For those accesses that can be statically verified, the runtime security check will also be eliminated.

{\bf Write-only Check.}
By default, \sysname checks all pointer dereferences, no matter read or write access.
In practice, a read access violation in general has much lower security implications, while a write access violation is the foundation of launching further exploits.
Thus, we provide an optional working mode of \sysname that focuses on integrity (but not confidentiality) and only enforces memory safety checks for memory write accesses, similar to existing sanitizers.
\section{Evaluation}
This section evaluates the \sysname prototype in terms of security and performance, and answers the following questions:

\begin{itemize}[noitemsep,nolistsep]
\item What types of vulnerabilities can \sysname detect?

\item Are there any types of vulnerabilities that \sysname can detect but other sanitizers cannot?

\item How much runtime performance and memory overheads does \sysname introduce?
\end{itemize}

\subsection{Environment and Comparison Targets}
\label{sec:env}
{\bf Environment.}
There are very few environments supporting ARM PA yet.
In addition to the official emulator ARM FVP~\cite{fvp}, only the new Apple M1 machines have support for ARM PA. But macOS does not provide ARM PA support for third-party applications yet.
Therefore, we build and run a Linux kernel v5.14.0 on a M1 mini device, and apply some custom patches to make it support more memory architectures.
On the other hand, we follow ARM's documentation~\cite{FVP-doc} to build the FVP test environment, and use the default memory configuration of FVP and apply it to the Linux kernel too, i.e., 39-bit for pointers and 24-bits for ARM PAC code. 



{\bf Comparison Targets.}
Moreover, we carefully choose several state-of-the-art (SOTA) open sourced solutions to conduct comparison experiments.
Among SOTA solutions, 
we only found HWASan and ASan integrated in the \texttt{clang} compiler support the ARM architecture\footnote{Another solution PARTS~\cite{parts} provides pointer integrity guarantee rather than memory safety guarantee. So it is not a proper target to compare with.},
while most other sanitizers, e.g., LowFat~\cite{lowfat-stack}, Memcheck~\cite{nethercote2007valgrind,seward2005using}, and SoftBound~\cite{softbound} and CETS~\cite{cets}, only support the x86 architecture.
So, we evaluate the security and performance of \sysname, ASan, and HWASan on the ARM architecture,
and evaluate that of ASan, LowFat, Memcheck, and SoftBound+CETS on the x86 architecture,
then use the evaluation results of ASan as a bridge to compare \sysname with other x86-based sanitizers.

\begin{table}[t]
  \scriptsize
    \centering
  \caption{Comparison of the time overhead of PA-related instructions running on M1 and FVP respectively.}
  \setlength{\tabcolsep}{1mm}{
\begin{tabular}{lllll}
\hline
                                                                                &                                                &        & Apple M1 & ARM FVP    \\ \hline
\multirow{6}{*}{\begin{tabular}[c]{@{}l@{}}ARM PA* \\ Instructions\end{tabular}} & \multicolumn{1}{c}{\multirow{3}{*}{Signature}} & PACDA  & 2.34424 ns  & 10.7015 ns \\
                                                                                & \multicolumn{1}{c}{}                           & PACDZA & 2.3444 ns   & 10.7058 ns \\
                                                                                & \multicolumn{1}{c}{}                           & PACGA  & 2.34435 ns  & 10.7012 ns \\ \cline{2-5} 
                                                                                & \multirow{2}{*}{Authentication}                & AUTDA  & 2.34465 ns  & 10.7016 ns \\
                                                                                &                                                & AUTDZA & 2.34465 ns  & 10.7016 ns \\ \cline{2-5} 
                                                                                & Stripping                                      & XPACD  & 2.34464 ns  & 10.7019 ns \\ \hline
\multicolumn{2}{l}{\multirow{2}{*}{ALU Operations}}                                                                              & ADD    & 0.334955 ns & 10.7013 ns \\
\multicolumn{2}{l}{}                                                                                                             & AND    & 0.334946 ns & 10.7014 ns \\ \hline
\multicolumn{3}{l}{PA Instructions : ALU Operations} & 7:1 & 1:1\\\hline
\end{tabular}
}
\label{tab:inst_cost}
\vspace{3mm}
\end{table}

\subsection{Microbenchmark Testing}

{\bf Performance gap between emulator and real devices.}
We have evaluated the performance of PA instructions on both the ARM FVP and Apple M1 mini respectively.
The results are listed in Table~\ref{tab:inst_cost}.
Although ARM FVP claims to be an accurate behavior model of ARM architecture~\cite{fvp},
we found that there is a significant performance gap.
Specifically, a PA instruction roughly takes 7x time as an ALU operation on real devices, while in ARM FVP the cost is the same.
Therefore, previous ARM PA work~\cite{parts, pacstack} based on PA-analogue~\cite{parts} have inaccurate performance results.
PA-analogue uses four exclusive-or (eor) operations to measure overhead of a PA instruction, while in reality the cost is much higher.
\textit{To the best of our knowledge, we are the first to evaluate the overheads of ARM PA-based solutions on real hardware.}

\begin{table}[t]
  \scriptsize
  \caption{Number of live objects in long-running programs. For Nginx, Apache, and Node.js, we used ApacheBench for stress testing, making 10,000 requests with 2,000 concurrent threads. For Redis, we used Redis-benchmark for 1,000,000 requests with 3,000 concurrent threads.}
  \centering
  \begin{tabular}{@{}ccr@{}}
    \toprule
    \multicolumn{2}{c}{Programs}             & Maximum number of objects \\ \midrule
    \multicolumn{2}{c}{Nginx}                & 5,635                      \\
    \multicolumn{2}{c}{Apache}               & 7,886                      \\
    \multicolumn{2}{c}{Node.js}              & 89,873                     \\ \midrule
    \multirow{2}{*}{Redis} & Redis-server    & 7,544                      \\
                           & Redis-benchmark & 17,731                     \\ \bottomrule
    \end{tabular}
    \label{tab:count}
    \vspace{3mm}
  \end{table}
  
{\bf Whether the PAC length is sufficient for use?}
Since the PAC is 24-bits in our setting, then the metadata table has at most $2^{24}=16,777,216$ slots.
If a program has more objects than this number, then the metadata table will overfill and \sysname will have false positives. 
Therefore, we have evaluated the maximum number of live objects (including heap variables, global variables, and stack variables) in several long-running and large real world programs.
As shown in Table~\ref{tab:count}, they have less than 100 thousand live objects, which is far from the aforementioned maximum threshold.
\textit{Thus, we believe the PAC length is sufficient for use in practice.}


\subsection{Security Evaluation}
\subsubsection{\bf Test Suites of Vulnerabilities}

\begin{table}[b]
\scriptsize
\caption{The test sets selected in the {\tt Juliet} test suite.}
\centering
\setlength{\tabcolsep}{1mm}{
\begin{tabular}{cccc}
\toprule
Test Sets & Vulnerability Type          & All Cases &  Selected Cases\\ \midrule
CWE121    & Stack-based Buffer Overflow & 6200                 & 6104\\
CWE122    & Heap-based Buffer Overflow  & 7740                 & 7260\\
CWE124    & Buffer Underwrite           & 2336                 & 2240\\
CWE126    & Buffer Overread             & 1740                 & 1644\\
CWE127    & Buffer Underread            & 2336                 & 2240\\
CWE415    & Double Free                 & 1636                 & 1636\\
CWE416    & Use After Free              & 786                  & 786\\
CWE476    & NULL Pointer Dereference    & 612                  & 576\\
CWE761    & Invalid Free                & 576                  & 576\\ \bottomrule
\end{tabular}
}
\label{tab:vul-type}
\end{table}

\begin{table*}[htb]
\footnotesize
\caption{Security evaluation based on the {\tt Juliet} test suite running for 3 times. In each test set (of a specific vulnerability type), there are multiple {\tt GOOD} and {\tt BAD} test cases, as presented in the brackets.
Since SoftBound+CETS~\cite{cets,softbound} does not support C++ programs completely, they are evaluated on fewer test cases, as shown in the third to last column.}
\centering
\resizebox{0.88\textwidth}{18mm}{
\begin{tabular}{ccccccccccc|ccc}
\hline
                     & \multicolumn{2}{c}{\sysname}            & \multicolumn{2}{c}{HWASan}                  & \multicolumn{2}{c}{\begin{tabular}[c]{@{}c@{}}ASan\\ (without LSan)\end{tabular}} & \multicolumn{2}{c}{LowFat}            & \multicolumn{2}{c|}{Memcheck}         & \multicolumn{3}{c}{SoftBound+CETS}                        \\ \cline{2-14} 
Vulnerability        & FP & \cellcolor[HTML]{EFEFEF}FN       & FP       & \cellcolor[HTML]{EFEFEF}FN       & FP                       & \cellcolor[HTML]{EFEFEF}FN                             & FP & \cellcolor[HTML]{EFEFEF}FN       & FP & \cellcolor[HTML]{EFEFEF}FN       & \#CASES     & FP       & \cellcolor[HTML]{EFEFEF}FN       \\ \hline
CWE121 (3052+3052)   & 0  & \cellcolor[HTML]{EFEFEF}1.180\%  & 0        & \cellcolor[HTML]{EFEFEF}16.972\% & 0                        & \cellcolor[HTML]{EFEFEF}8.159\%                        & 0  & \cellcolor[HTML]{EFEFEF}46.888\% & 0  & \cellcolor[HTML]{EFEFEF}74.935\% & 2817+2817   & 2.520\%  & \cellcolor[HTML]{EFEFEF}25.240\% \\
CWE122 (3630+3630)   & 0  & \cellcolor[HTML]{EFEFEF}0.992\%  & 0        & \cellcolor[HTML]{EFEFEF}0.992\%  & 0                        & \cellcolor[HTML]{EFEFEF}2.314\%                        & 0  & \cellcolor[HTML]{EFEFEF}42.369\% & 0  & \cellcolor[HTML]{EFEFEF}27.053\% & 3331+3331   & 46.142\% & \cellcolor[HTML]{EFEFEF}13.990\%  \\
CWE124 (1120+1120)   & 0  & \cellcolor[HTML]{EFEFEF}0        & 0        & \cellcolor[HTML]{EFEFEF}7.202\%  & 0                        & \cellcolor[HTML]{EFEFEF}8.75\%                         & 0  & \cellcolor[HTML]{EFEFEF}0        & 0  & \cellcolor[HTML]{EFEFEF}34.375\% & 1030+1030   & 23.301\% & \cellcolor[HTML]{EFEFEF}10.971\% \\
CWE126 (822  +  822) & 0  & \cellcolor[HTML]{EFEFEF}0        & 0        & \cellcolor[HTML]{EFEFEF}6.650\%  & 0                        & \cellcolor[HTML]{EFEFEF}14.842\%                       & 0  & \cellcolor[HTML]{EFEFEF}12.409\% & 0  & \cellcolor[HTML]{EFEFEF}57.664\% & 760  +  760 & 19.079\% & \cellcolor[HTML]{EFEFEF}7.105\%  \\
CWE127 (1120+1120)   & 0  & \cellcolor[HTML]{EFEFEF}0        & 0        & \cellcolor[HTML]{EFEFEF}14.435\% & 0                        & \cellcolor[HTML]{EFEFEF}11.25\%                        & 0  & \cellcolor[HTML]{EFEFEF}14.911\% & 0  & \cellcolor[HTML]{EFEFEF}31.161\% & 1030+1030   & 23.301\% & \cellcolor[HTML]{EFEFEF}11.165\% \\
CWE415 (818  +  818) & 0  & \cellcolor[HTML]{EFEFEF}0        & 0        & \cellcolor[HTML]{EFEFEF}2.445\%  & 0                        & \cellcolor[HTML]{EFEFEF}0                              & 0  & \cellcolor[HTML]{EFEFEF}1        & 0  & \cellcolor[HTML]{EFEFEF}0        & 748  +  748 & 2.273\%  & \cellcolor[HTML]{EFEFEF}0        \\
CWE416 (393  +  393) & 0  & \cellcolor[HTML]{EFEFEF}0        & 0        & \cellcolor[HTML]{EFEFEF}17.303\% & 0                        & \cellcolor[HTML]{EFEFEF}0                              & 0  & \cellcolor[HTML]{EFEFEF}1        & 0  & \cellcolor[HTML]{EFEFEF}0        & 392  +  392 & 67.347\% & \cellcolor[HTML]{EFEFEF}10.204\% \\
CWE476 (288  +  288) & 0  & \cellcolor[HTML]{EFEFEF}0        & 0        & \cellcolor[HTML]{EFEFEF}0        & 0                        & \cellcolor[HTML]{EFEFEF}0                              & 0  & \cellcolor[HTML]{EFEFEF}0        & 0  & \cellcolor[HTML]{EFEFEF}0        & 270  +  270 & 16.296\% & \cellcolor[HTML]{EFEFEF}0        \\
CWE761 (288  + 288)  & 0  & \cellcolor[HTML]{EFEFEF}0        & 0        & \cellcolor[HTML]{EFEFEF}1        & 0                        & \cellcolor[HTML]{EFEFEF}0                              & 0  & \cellcolor[HTML]{EFEFEF}1        & 0  & \cellcolor[HTML]{EFEFEF}0        & 264  +  264 & 2.273\%  & \cellcolor[HTML]{EFEFEF}0        \\ \hline
\end{tabular}
}
\label{tab:security_evaluate}
 \vspace{-2mm}
\end{table*}

{\tt Magma}~\cite{Magma} provides a benchmark of ground truth security-critical ground truth
bugs in commonly-used open-source libraries.
We used the {\tt Magma} test suit to evaluate the detection capability of \sysname and ASan on real-world programs.

To further systematically evaluate the security and functionality correctness of \sysname,
we utilize test sets from the {\tt Juliet} test suite consisting of many memory safety vulnerabilities to conduct experiments.
Among those test sets, we have selected the test sets pertaining to spatial and temporal memory safety.
These selected test sets are listed in Table~\ref{tab:vul-type}.

As listed in Table~\ref{tab:vul-type}, 
a small number of test cases are not selected to test, because they cannot trigger the intended vulnerabilities, in order to measure the false-negative rates accurately.
Some of them have no vulnerabilities at runtime, but are included in the test suite for evaluating the accuracy of static analysis tools.
Some others cannot trigger the vulnerability because of the dependency of architecture  or random number.
Details of such cases are listed in Appendix~\ref{sec:appendix-juliet}.

\subsubsection{\bf Metrics and Configuration}
To evaluate the security performance, we select five classic and commonly-used open-source sanitizers, i.e., HWASan, ASan, LowFat, Memcheck, SoftBound+CETS, and compare their performance with \sysname.

{\bf Metric.}
For each test case, the {\tt Juliet} test suite generates two test programs, \texttt{GOOD} and \texttt{BAD}.
Each \texttt{GOOD} program is not vulnerable, and each \texttt{BAD} program has an exploited memory corruption.
Therefore, it is a false positive for a sanitizer to report a \texttt{GOOD} program as anomalous.
Also, it is a false negative for a sanitizer not to report a \texttt{BAD} as anomalous.

{\bf Sanitizer configuration.}
To provide a fair comparison,
we made every effort to reduce the false negatives and false positives of other sanitizers caused by implementations. 
For example, in some test cases of the {\tt Juliet}, 
some memory objects are used without proper deallocation and therefore detected by the LeakSanitizer embedded in ASan, which causes ASan to report a false positive. 
The same situation exists in Memcheck. 
So we disable the memory leak detection mechanism of both ASan and Memcheck. 

In addition, LowFat may put objects in aligned but larger memory blocks (e.g., a 10-bytes object takes 16 bytes memory).
As a result, an illegal input that causes out-of-bound access (e.g., accessing 11 to 16 bytes) may get silently mitigated, and LowFat will not report an alarm.
To avoid this type of unintended false negatives, we increase the input values of {\tt BAD} programs in the {\tt Juliet} test suite,  enabling LowFat to report an alarm when such objects are overflowed.


\begin{table}[htb]
\scriptsize
\caption{Number of proof-of-concept (poc) samples provided by Magma (from the AFL++ directory), the number of PoCs  that \sysname detected violations, and the number of PoCs that ASan detected violations.}
\centering
\setlength{\tabcolsep}{1mm}{
\begin{tabular}{cccc}
\toprule
open-source libraries           & PoCs  & \sysname        & ASan          \\ \midrule
libpng     & 634   & 0             & 0             \\
LibTIFF    & 3716  & 115           & 115           \\
Libxml2    & 19614 & 0             & 0             \\
Poppler    & 7343  & 1329          & 1329          \\
OpenSSL    & 655   & 194           & 194           \\
SQLite     & 1777  & 0             & 0             \\
PHP        & 1443  & \textbf{1128} & \textbf{1083} \\
Lua        & 0     & 0             & 0             \\
libsndfile & 0     & 0             & 0             \\ \bottomrule
\end{tabular}
}
\label{tab:magma}
\end{table}

\subsubsection{\bf Results of Security Evaluation}
{\bf \sysname has better detection capability than ASan on the real-world bugs.}
Proof of concept (PoC) exploits are used to reveal the security weaknesses within software.
Magma~\cite{Magma} collects some open-source libraries with real-world vulnerabilities and provides sufficient PoCs.
We used PoCs provided by Magma to evaluate target programs instrumented by \sysname or ASan, respectively.
As we can see from Table~\ref{tab:magma}, \sysname and Asan detect the same number of PoCs for most programs.
For PHP, \sysname found 45 more PoCs than ASan, all of which were PoCs from CVE-2018-14883, a heap-based buffer over-read caused by an integer overflow. 
ASan fails to detect 45 PoCs because the illegal memory access crosses the RedZone instrumented by ASan and access another valid object.
However, since \sysname can accurately obtain the base address and the object size of the memory object to which the pointer corresponds, \sysname can effectively detect these PoCs.
We have analyzed the results in detail in Appendix~\ref{sec:appendix-3} and provided detail results of {\tt Magma}.
Note that, Magma contains not only just memory corruption vulnerabilities, and the PoCs may not trigger all the vulnerabilities. So it is expected that some target programs do not detect an exception once. The evaluation focuses on the comparison with the common-used ASan.

{\bf \sysname has no false positives.}
The results of the Juliet test suit are shown in Table~\ref{tab:security_evaluate}.
\texttt{FP} represents the False-Positive rate,  and \texttt{FN} represents the False-Negative rate. 
As mentioned before, \sysname has no false positives because it uses rehashing to avoid the only source of positives -- collision of two different active metadata (i.e., taking a same table slot).

%

%
{\bf \sysname provides better memory safety guarantees.}
While ASan has the lowest false-negative rates among existing open-source sterilizers, \sysname has even fewer false negatives than ASan, thus provides better memory safety guarantees.
\sysname cannot detect sub-object overflow, which is a common problem of all these sanitizers, including ASan, HWASan, LowFat, and Memcheck, due to the trade-off with metadata tracking overhead.
SoftBound claims to detect sub-object overflow, but it does not detect any of the sub-object overflow test cases in the {\tt Juliet} test, as we will discuss the root cause later.
Additionally, \sysname covers more types of vulnerabilities than HWASan and LowFat.

{\bf Summary}\quad
Overall, \sysname has better detection ability than ASan, HWASan, LowFat, Memcheck, and SoftBound+CETS.
\sysname performs well in detecting spatial and temporal memory vulnerabilities.
Specifically, \sysname performs the authentication for the pointer passed to the \texttt{free} function before deallocating a memory object, so \sysname can effectively detect double free and invalid free bugs.
Since the memory safety check is required for each memory operation, \sysname can also detect out-of-bounds access, use-after-free, and null pointer dereference.

\subsubsection{\bf Root Cause Analysis of False Negatives}
In general, existing sanitizers have a much larger false negatives, i.e., they would miss real vulnerabilities at runtime.
The reasons for the false negatives of these sanitizers are analyzed and discussed in detail in the following paragraphs, both in terms of implementation and design.

{\bf \sysname}\quad
The false negatives for \sysname are due to test cases with sub-object overflow.
Since the metadata of \sysname just contains the size of the entire object and has no information about the sub-objects in the target object, \sysname cannot detect such violations.

{\bf HWASan}\quad
In terms of design, on the one hand, HWASan only matches tags on memory accesses and therefore cannot detect dangling pointers passed to third-party libraries; 
on the other hand, if a pointer passed to the \texttt{free} function does not point to the beginning of an object, HWASan cannot detect it because its tag does match the tag of the object.
In other words, HWASan cannot detect invalid free bugs.
Besides, HWASan also cannot detect sub-object overflow.
Moreover, HWASan relies on 8-bits tags and, in practice, many tags conflicted. 
We ran {\tt CWE415} three times and found that the conflicting tags caused false negatives for some cases. 
Therefore, false negatives for HWASan are the average of the three results.
In terms of implementation, HWASan cannot handle stack variables allocated by the {\tt ALLOCA} function, 
so no buffer overflows of such memory objects can be detected by HWASan. 
In addition, HWASan does not handle some common memory access functions such as the {\tt snprintf} and {\tt strncat} functions.

{\bf ASan.}\quad
In terms of design, ASan cannot detect cross-object overflow that skips redzones, or access an object that has been freed for a while, as shown in MEDS~\cite{meds}.
But \sysname can effectively detect such cases.
However, \sysname and ASan both cannot detect sub-object overflow.
In terms of implementation, we found that the function {\tt \_\_asan\_alloca\_poison} used by ASan to set the redzones does not work in some cases,  causing false negatives.

{\bf LowFat}\quad
LowFat, a sanitizer for detecting spatial memory vulnerabilities, cannot detect any temporal memory vulnerabilities.
In terms of design, a reason for the high false-negative rates in {\tt CWE121} is that some test cases use indexes to access arrays. 
LowFat uses the pointer itself to find the corresponding metadata, so it ignores such cases that a memory operation using pointer offsets accesses another memory object.
In such cases, the access length does not exceed the boundary of the victim object, LowFat considers the operation as legitimate.
Array access is the most common cause of such cases, and we provide cases for further analysis and discussion in the Appendix~\ref{sec:appendix-2}.
Moreover, LowFat also cannot detect sub-object overflow.
In terms of implementation, since LowFat rounds up the allocated size of an object to a proper number to facilitate metadata (i.e., the bounds) retrieval, 
it cannot detect memory access that is beyond the object size but within the allocated size, 
which is a reason for high false-negative rates in some test sets.
Besides, similar to HWASan, LowFat does not handle the {\tt snprintf} functions.

{\bf Memcheck}\quad
In terms of design, Memcheck has a high false-negative rate in some test sets due to its inability to detect memory overflows in global or local objects.
Apparently, Memcheck also cannot detect sub-object overflow.

{\bf SoftBound+CETS}\quad \label{SoftBound}
SoftBound+CETS~\cite{cets,softbound} is believed to be a good combination providing full memory safety.
However, it has a much higher false positive rate and false negative rate.
Regarding false negatives, likely HWASan,  SoftBound+CETS does not handle some common memory access functions such as the {\tt strncat} functions and it also cannot detect the dangling pointers passed to third-party libraries.
Regarding false positives, there are some flaws in the prototype implementation of SoftBound+CETS, and thus it has a high false positive rate.
Besides, it currently cannot support C++ programs, reducing the number of supported test cases.
Moreover, even though SoftBound claims it can detect sub-object overflow in
theory, the security tests have shown that the implementation cannot.
We extensively analyze the current prototype using the intermediate LLVM bitcode.
If the overflowing sub-object is the first sub-object of a structure, then it uses the boundary information of this structure for the bound check.
The SoftBound paper does not elaborate on distinguishing boundary information when the program uses an object pointer to load the bound information,
and the implementation falls back to the first sub-object to access the memory.
Specifically, SoftBound uses the addresses in pointers to retrieve their bound information and cannot distinguish an object pointer and the first sub-object pointer of this object,
who have the same addresses but different bounds.
While the approach works for some cases, it is not generalizable.
Take {\tt *char\_type\_overrun\_memcpy\_01} in {\tt CWE121} for example, where the {\tt memcpy} function is called to assign value to a 16-byte array,  i.e. the first sub-object of the global value `charVoid`.
In the {\tt bad} case, the {\tt memcpy} function uses the size of global value `charVoid` instead of 16 bytes. 
SoftBound calls {\tt \_\_softboundcets\_memcpy\_check} before calling the
{\tt memcpy} function, which uses the bounds information of the global value `charVoid` rather than that of the first sub-object, 
so SoftBound cannot detect the sub-object overflow.

\subsubsection{\bf Comparison between Sanitizers}

\sysname outperforms other sanitizers in detecting out-of-bounds access, UAF, and invalid free vulnerabilities.

\begin{itemize}[noitemsep,nolistsep]
\item For out-of-bounds access, if attackers can control the index used to access an array object to access another valid object out-of-bounds, ASan, Memcheck, and LowFat cannot detect such out-of-bounds access.
\item For UAF, if the quarantine used by Asan is exhausted, Asan cannot detect UAF effectively; Memcheck cannot detect such UAF cases that the freed object is reallocated.
\item For invalid free, since the pointer used to be freed still points to the middle of the object, its tag still matches the object's tag and can pass the HWASan security checks. So HWASan cannot detect invalid free vulnerabilities.
\end{itemize}

The above vulnerabilities can be detected by \sysname effectively.
In Appendix~\ref{sec:appendix-2}, we will use specific examples to demonstrate the advantages of \sysname.

\begin{table*}[]
  \footnotesize
  \caption{Runtime overheads of \sysname and \sysname write-only, compared with HWASan, ASan, LowFat, Memchck, SoftBound+MPX, and SoftBound+CETS. Since LowFat has an outlier, the average overhead when including the outlier and excluding the outlier are calculated separately.} 
  \centering
  \resizebox{0.9\textwidth}{!}{
  \begin{tabular}{
    >{\columncolor[HTML]{FFFFFF}}l 
    >{\columncolor[HTML]{FFFFFF}}c 
    >{\columncolor[HTML]{FFFFFF}}c 
    >{\columncolor[HTML]{FFFFFF}}c 
    >{\columncolor[HTML]{FFFFFF}}c 
    >{\columncolor[HTML]{FFFFFF}}c 
    >{\columncolor[HTML]{FFFFFF}}c 
    >{\columncolor[HTML]{FFFFFF}}c 
    >{\columncolor[HTML]{FFFFFF}}c 
    >{\columncolor[HTML]{FFFFFF}}c 
    >{\columncolor[HTML]{FFFFFF}}c }
    \hline
    \multicolumn{2}{c}{\cellcolor[HTML]{FFFFFF}}                                          & \multicolumn{4}{c}{\cellcolor[HTML]{FFFFFF}AArch64}                                  & \multicolumn{5}{c}{\cellcolor[HTML]{FFFFFF}X86}                         \\ \cline{3-11} 
    \multicolumn{2}{c}{\multirow{-2}{*}{\cellcolor[HTML]{FFFFFF}Test Case}}               & \sysname & \sysname(+w.o) & ASAN   & \multicolumn{1}{c|}{\cellcolor[HTML]{FFFFFF}HWASAN} & ASAN        & LOWFAT         & Memcheck & SoftBound+CETS & SoftBound+MPX \\ \hline
    \multicolumn{2}{l}{\cellcolor[HTML]{FFFFFF}600.perlbench\_s}                          & 2.0083 & 1.0163       & 1.5910 & \multicolumn{1}{c|}{\cellcolor[HTML]{FFFFFF}3.6079} & 1.6362      & 206.4738       & 35.2600  & $\varnothing$            & $\varnothing$           \\
    \multicolumn{2}{l}{\cellcolor[HTML]{FFFFFF}602.gcc\_s}                                & 2.0127 & 1.2025       & 1.3544 & \multicolumn{1}{c|}{\cellcolor[HTML]{FFFFFF}2.7031} & 1.1414      & 1.8870         & 23.3416  & $\varnothing$            & $\varnothing$           \\
    \multicolumn{2}{l}{\cellcolor[HTML]{FFFFFF}605.mcf\_s}                                & 1.0720 & 0.5769       & 0.5282 & \multicolumn{1}{c|}{\cellcolor[HTML]{FFFFFF}0.3613} & 1.2853      & 0.6283         & 18.40636 & $\varnothing$           & $\varnothing$          \\
    \multicolumn{2}{l}{\cellcolor[HTML]{FFFFFF}619.lbm\_s}                                & 0.0290 & 0.0169       & 0.1097 & \multicolumn{1}{c|}{\cellcolor[HTML]{FFFFFF}0.1445} & 0.7502      & -0.0367        & 3.1156   & -0.2606       & -0.5250       \\
    \multicolumn{2}{l}{\cellcolor[HTML]{FFFFFF}625.x264\_s}                               & 0.0555 & -0.1143      & 0.8818 & \multicolumn{1}{c|}{\cellcolor[HTML]{FFFFFF}1.0580} & 1.5173      & 3.7046         & 36.7384  & $\varnothing$            & $\varnothing$          \\
    \multicolumn{2}{l}{\cellcolor[HTML]{FFFFFF}638.imagick\_s}                            & 0.9578 & 0.3008       & 1.5480 & \multicolumn{1}{c|}{\cellcolor[HTML]{FFFFFF}0.8062} & 0.6953      & 0.7994         & 21.2429  & $\varnothing$            & -0.1961           \\
    \multicolumn{2}{l}{\cellcolor[HTML]{FFFFFF}644.nab\_s}                                & 1.1293 & 0.2810       & 1.3151 & \multicolumn{1}{c|}{\cellcolor[HTML]{FFFFFF}1.1694} & 0.0444      & 0.2165         & 6.8393   & -0.02196      & -0.3817         \\
    \multicolumn{2}{l}{\cellcolor[HTML]{FFFFFF}657.xz\_s}                                 & 0.1575 & 0.1028       & 0.0993 & \multicolumn{1}{c|}{\cellcolor[HTML]{FFFFFF}0.1887} & 0.6206      & 0.7278         & 17.9604  & $\varnothing$            & $\varnothing$           \\
    \multicolumn{2}{l}{\cellcolor[HTML]{FFFFFF}nginx}                                     & 0.1265 & 0.0374       & 0.7043 & \multicolumn{1}{c|}{\cellcolor[HTML]{FFFFFF}1.9084} & 0.6541      & $\ast$          & $\ast$    & $\ast$          & 4.9317        \\ \hline
    \multicolumn{2}{l}{\cellcolor[HTML]{FFFFFF}Average}                                   & {\bf 0.8387} & {\bf 0.3800}       & 0.9035 & \multicolumn{1}{c|}{\cellcolor[HTML]{FFFFFF}1.3275} & 0.9272     & 26.8001/1.1324 & 20.3631  & -0.14128      & 0.957225   \\ \hline
    \end{tabular}
    \label{tab:performance}
  }
   \begin{tablenotes}
    \item $\varnothing$: Compilation failed. $\ast$: Crashed at runtime
  \end{tablenotes}
  \end{table*}

\begin{table*}[]
  \footnotesize
    \caption{Memory usage of \sysname, HWASan, ASan, LowFat, Memchck, SoftBound+MPX, and SoftBound+CETS. Since LowFat has an outlier, the average overhead when including the outlier and excluding the outlier are calculated separately.}
\centering
\resizebox{0.9\textwidth}{!}{
  \begin{tabular}{
    >{\columncolor[HTML]{FFFFFF}}l 
    >{\columncolor[HTML]{FFFFFF}}l 
    >{\columncolor[HTML]{FFFFFF}}c 
    >{\columncolor[HTML]{FFFFFF}}c 
    >{\columncolor[HTML]{FFFFFF}}c 
    >{\columncolor[HTML]{FFFFFF}}c 
    >{\columncolor[HTML]{FFFFFF}}c |
    >{\columncolor[HTML]{FFFFFF}}c 
    >{\columncolor[HTML]{FFFFFF}}c 
    >{\columncolor[HTML]{FFFFFF}}c 
    >{\columncolor[HTML]{FFFFFF}}c 
    >{\columncolor[HTML]{FFFFFF}}c 
    >{\columncolor[HTML]{FFFFFF}}c }
    \hline
    \multicolumn{2}{l}{\cellcolor[HTML]{FFFFFF}}                            &         & \multicolumn{4}{c}{\cellcolor[HTML]{FFFFFF}AArch64} &          & \multicolumn{5}{c}{\cellcolor[HTML]{FFFFFF}X86}                                    \\ \cline{3-13} 
    \multicolumn{2}{l}{\multirow{-2}{*}{\cellcolor[HTML]{FFFFFF}Test Case}} & Origin  & \sysname         & \sysname(+w.o)   & ASAN    & HWASAN  & Origin   & ASAN    & LOWFAT      & Memcheck & SoftBound+CETS                  & SoftBound+MPX \\ \hline
    \multicolumn{2}{l}{\cellcolor[HTML]{FFFFFF}600.perlbench\_s}            & 158296 KB  & 1.6794         & 1.6673         & 6.0813  & 0.6972  & 158976   & 4.3550  & 1174.2286   & 1.8150   & $\varnothing$                             & $\varnothing$           \\
    \multicolumn{2}{l}{\cellcolor[HTML]{FFFFFF}602.gcc\_s}                  & 95464 KB  & 3.3120         & 3.2327         & 4.5541 & 0.8374  & 96924    & 4.0906  & 0.2690      & 1.8316   & $\varnothing$                             & $\varnothing$           \\
    \multicolumn{2}{l}{\cellcolor[HTML]{FFFFFF}605.mcf\_s}                  & 558784 KB  & 0.0901        & 0.0895         & 0.5288  & 0.0723  & 559024   & 0.5254  & 0.2283      & 0.3969   & $\varnothing$                             & $\varnothing$           \\
    \multicolumn{2}{l}{\cellcolor[HTML]{FFFFFF}619.lbm\_s}                  & 3302496 KB & -0.0001       & -0.0001         & 0.0692  & 0.06291  & 3305228  & 0.1271  & 0.0004      & 0.3406   & -0.0008                         & -0.0008       \\
    \multicolumn{2}{l}{\cellcolor[HTML]{FFFFFF}625.x264\_s}                 & 120038 KB   & 0.1145      & 0.1094         & 0.3621  & 0.1507  & 120107   & 0.2498  & 6.4748      & 0.9560   & $\varnothing$                             & $\varnothing$           \\
    \multicolumn{2}{l}{\cellcolor[HTML]{FFFFFF}638.imagick\_s}              & 22500 KB   & 0.3710         & 0.3429         & 1.4601  & 0.0030 & 16436    & 2.0402  & -0.0187     & 17.5378  & $\varnothing$                             & 0.1414           \\
    \multicolumn{2}{l}{\cellcolor[HTML]{FFFFFF}644.nab\_s}                  & 11540 KB   & 9.2759         & 9.2561         & 70.6009 & 3.7266  & 16658    & 40.8168 & 0.0725      & 25.7013  & -0.0230                            & -0.3817           \\
    \multicolumn{2}{l}{\cellcolor[HTML]{FFFFFF}657.xz\_s}                   & 4529600 KB & 0.0003        & 0.0003        & 0.0459  & 0.0666  & 17291512 & 0.0158  & -0.0009     & 0.2677   & $\varnothing$                             & $\varnothing$           \\
    \multicolumn{2}{l}{\cellcolor[HTML]{FFFFFF}nginx}                       & 3256 KB   & 2.4361         & 2.1916         & 74.2887   & 4.0418  & 1884     & 3.8089  & $\ast$       & $\ast$    & $\ast$                           & 28.6157       \\\hline
    \multicolumn{2}{l}{\cellcolor[HTML]{FFFFFF}Average}                     &         & \textbf{1.9199}  & \textbf{1.8766}  & 17.5546 & 1.0732   &          & 6.2255   & 147.6568 / 1.0036 & 6.1059    & \cellcolor[HTML]{FFFFFF}-0.0119 & 7.09365        \\ \hline
    \end{tabular}
}\label{table-memory}
 \begin{tablenotes}
    \item $\varnothing$: Compilation failed. $\ast$: Crashed at runtime
  \end{tablenotes}
\end{table*}


\subsection{Performance Evaluation}
{\bf Test suite and inputs.}
Regarding the performance evaluation,
we use all C benchmarks from the SPEC CPU2017 (using the SPECspeed 2017 suite) and Nginx to evaluate the performance overheads.
Four test cases do not work correctly with the reference input on some sanitizers we compare with.
In order to conduct a fair comparison, only these targets are changed
to utilize the train input (which is more complex than the test input), and the rest of the targets use the reference input. 
Specifically, the reference input of 602 causes a segmentation fault in LowFat; 
the reference input of 605 causes HWASan to report the error: ``Unable to allocate memory''; 
the reference inputs of 638 and 644 cause MemCheck to enter an endless loop
(running for 1$\sim$3 days without stopping).
It is worth noting that all security tests and performance tests were run on real hardware, i.e., the Apple M1 mini.

{\bf Experimental Setup.}
\sysname, ASan, and HWASan all detect global variable overflow for the {\tt ldecoder} of {\tt x264\_s} benchmark and {\tt gcc\_s} benchmark.
We manually confirmed that these are not false alarms.
But, LowFat indeed has some false positives. 
However, we need to run the program completely to measure the performance overheads.
Thus, we modify the default configuration to allow the program to run normally until completion, even if the sanitizer detects some violation.

{\bf Runtime overhead.}
Table~\ref{tab:performance} shows the runtime overhead for \sysname, HWASan, and ASan on the ARM architecture.
The average runtime overheads for ASan and HWASan are 0.904$\times$ and 1.328$\times$, respectively.
And the average runtime overhead of \sysname is 0.84$\times$, which is lower than ASan and HWASan (while providing stronger security guarantees).
We also tested the runtime overhead of the weaker form of \sysname, which only enforces memory safety checks for memory write accesses. 
The results are shown in Table~\ref{tab:performance} as well (denoted as \sysname w.o.). 
The runtime overhead of \sysname w.o. is 0.38$\times$, which is much lower than others.

Other open-source sanitizers do not support the ARM architecture, but we still performed runtime tests on the x86 architecture, and the results are shown in Table~\ref{tab:performance}.
Compared to other open-source sanitizers on x86 architecture, ASan has the lowest runtime overhead.
LowFat has an outlier when running {\tt perlbench\_s}, but the average runtime overhead after subtracting this outlier is still higher than ASan.
Through the above comparison, we believe that \sysname will have a significant runtime overhead advantage if these sanitizers can be evaluated on the same experimental platform.

{\bf Memory Overhead.}
The memory usage of all sanitizers is shown in Table~\ref{table-memory}.
The memory consumption of \sysname is much lower than that of ASan, and Memcheck and SoftBound+MPX.
Although the memory consumption of \sysname is higher than HWASan and LowFat, LowFat only detect spatial safety violations and the runtime overhead of HWASan is higher than that of \sysname.

{\bf Summary}\quad
Overall, \sysname has lower runtime overhead than HWASan and ASan.
And the memory overhead of \sysname has significantly lower than ASan but higher than HWASan. However, \sysname is significantly better at security capability than HWASan.
Specifically, \sysname has reduced \runperf of runtime overheads and \memperf of memory overheads than ASan.
\textit{Considering all performance overheads and security detection capabilities, \sysname is a better choice for balancing performance and security.}

 \section{Limitation}\label{sec:limitation}
{\bf Limited PACs.}
Table~\ref{tab:count} demonstrates that the 24-bit PA length is sufficient for regular programs. 
If the metadata table needs to grow, 
\sysname can use techniques like linked lists, 
to resolve collision issues when the number of PACs exceeds its upper limit (e.g., if programs require more than 16,777,216 objects that are live at the same time).

{\bf Sub-object Overflow.}
Same as many other common sanitizers (such as AddressSanitizer~\cite{asan}, LowFat~\cite{lowfat-heap,lowfat-stack} and BaggyBounds~\cite{akritidis2009baggy}),
\sysname cannot detect overflows within object, i.e., sub-object overflow.
SoftBound~\cite{softbound} uses static type information
(i.e., from the source code) to narrow bounds to detect sub-object overflow in
nearby location. This best-effort approach
requires the object fields' metadata to be associated with pointers.
In its current implementation,
SoftBound cannot detect sub-object overflows if casting is involved and
LLVM optimizations may also cause SoftBound to miss sub-object overflows.
Besides, SoftBound cannot detect the case discussed in Section~\ref{SoftBound}.
It is hard to track object fields' metadata effectively, and it always causes very high overheads.
We therefore leave it as future work.

\section{Related Work}
\subsection{Memory Safety sanitizer}
Security researchers use sanitizers to detect security vulnerabilities dynamically.
Different sanitizers target different classes of vulnerabilities, e.g., spatial safety violations~\cite{kuvaiskii2017sgxbounds,softbound,lowfat-heap,lowfat-stack}, temporal safety violations~\cite{DangNull,dangsan2017,Undangle,FreeSentry}, type confusion~\cite{typesan, jeon17ccs}, or undefined behavior~\cite{ubsan}.
Some sanitizers~\cite{dangsan2017,DangNull,jones1997backwards,ruwase2004practical,dhurjati2006backwards} only focus on spatial safety violations or temporal safety violations. 
We further discuss some classical and commonly-used sanitizers.

Memcheck~\cite{nethercote2007valgrind,seward2005using} is a memory corruption detector. 
It maintains a valid-address table to determine whether a pointer being dereferenced is valid and maintains a valid-value list to check whether the accessed object has been initialized.
However, Memcheck cannot detect memory overflow in global variables and stack variables and cannot detect UAF since the freed object could be taken by another valid and initialized object.
Besides, Memcheck introduces excessive-performance overheads.
SoftBound+CETS~\cite{softbound,cets} provides a full memory safety solution.
SoftBound+CETS utilizes the pointer-based bounds-checking and identifier metadata associated with pointers. 
However, SoftBound+CETS introduces high runtime overheads.
AddressSanitizer (ASan)~\cite{asan} has a better performance which utilizes the redzones to detect buffer overflows and uses a quarantine to catch Use-After-Free vulnerabilities. 
Utilizing the redzones can quickly detect memory vulnerabilities, but the redzones introduce high memory overheads.
Besides, ASan cannot effectively detect UAF vulnerabilities after the quarantine has been exhausted.
LowFat~\cite{lowfat-heap,lowfat-stack} further reduces overhead, which utilizes the bound check to detect out-of-bounds errors.
It uses a special encoding scheme to encode the bounds into the pointers themselves for fast retrieval.
However, LowFat just can detect spatial memory safety vulnerabilities.
And if a malicious pointer accesses another object beyond the bounds and the access length does not exceed the bounds of the victim object, LowFat cannot effectively detect such spatial safety violations.


\subsection {\bf Metadata Management}\label{sec:metadata}
Sanitizers, in general, will use metadata (such as bound information or tags) to track memory safety states and catch memory access violations at runtime.
 
ASan~\cite{asan} and HWAsan~\cite{HWASAN} use the direct-mapped shadow to store the metadata for a block of 8 bytes.
It is very efficient, but it wastes the memory occupied by metadata. 
TypeSan~\cite{typesan}, Intel MPX~\cite{mpx}, and METAlloc~\cite{haller2016metalloc} use a multi-level shadow to index metadata. 
Compared with the direct-mapped shadow, this method can effectively reduce the memory of metadata, but significantly degrades the runtime performance of programs with frequent memory accesses.
Oscar~\cite{dang2017oscar} and PTAuth~\cite{ptauth} increase the allocation size of each object and append the metadata to the data of the objects. 
When the object's size is large enough, this approach may have false positives or excessive performance overheads.
Many efforts~\cite{watchdog,hardbound,woodruff2014cheri,woodruff2019cheri} embed metadata into pointers and implement different forms of fat pointers by extending registers or language implementations. 
However, the approach needs to change the processor hardware and increases runtime overheads.
LowFat~\cite{lowfat-heap,lowfat-stack} encodes the object-bound information into the pointer itself via a special encoding scheme and has good compatibility. 
However, it modifies the heap allocator to round up the allocated sizes of objects to facilitate metadata (i.e., the bounds) retrieval, 
and stores objects with different sizes into different memory pages, 
causing significant memory fragmentation problems.
Some mechanisms~\cite{cets,Undangle} use disjoint metadata to improves the compatibility. 
The disjoint metadata often corresponds to each pointer, 
so during the pointer propagation process, it is necessary to maintain the propagation and copy of its metadata.

\subsection{\bf Hardware Expansions}
HWASan~\cite{HWASAN} and ARM MTE~\cite{MTE} use the higher-order bits of the pointer to store the tag, and memory access is only possible if the tag of the pointer and the memory are identical.
The memory tag in HWASan is 8 bits, and ARM MTE is a 4-bit integer associated with each 16-byte aligned memory region.
Due to the length of the tag, the probability of tag collision is 6.25\%.
So MTE and HWASan can only probabilistically detect memory corruption.

AOS~\cite{aos} is a heap memory safety guard. 
This scheme implements a set of variant instructions based on PA and some additionally needed hardware extensions.
\sysname and AOS also use tags to index metadata, but there are critical differences between these two schemes. 
On the one hand, AOS uses one tag for multiple metadata sets to resolve hash collisions, 
which perhaps increases the number of queries at the time of visit. 
When the metadata table is full, 
AOS needs to increase the number of sets for each tag to extend the metadata table, 
which is expensive because of the copy of all the entries.
\sysname ensures the uniqueness of tags during tag generation, thus avoiding hash collisions.
On the other hand, AOS uses SP to generate the pointer signature, uses the object size to generate the 2-bits address hashing code (AHC), and may fail to detect use-after-realloc vulnerabilities (with the same base address, object size, and SP). 
\sysname uses SP and a static random number to yield a random and conflict-free birthmark, which further contributes to the pointer authentication code, and thus avoids such FNs. 
In addition, this scheme only strips pointers when memory objects are deallocated and does not verify the integrity of pointers, so it cannot detect invalid free vulnerabilities.
Moreover, AOS requires changes to the processor, ISA, and the OS kernel. These extra required hardware extensions increase hardware costs and make its adoption challenging. 
However, \sysname takes advantage of existing hardware features without any hardware modifications.
Compared with AOS, \sysname is more practical (compared to AOS and does not require hardware modifications), secure, and efficient.

BOGO~\cite{zhang2019bogo}  utilizes  Intel MPX to implement the spatial and temporal security.
However, Intel claims that MPX has been deprecated and both GCC and Linux kernel no longer support Intel MPX.
The method of No-FAT~\cite{ziad2021no} indexing metadata is similar to LowFat in that it uses the pointer itself to compute the object's base address and thus check if memory accesses are out of bounds. 
The difference with LowFat is that No-Fat's base address calculation and memory access checking are implemented in hardware.
Furthermore, No-Fat also uses tags to catch temporal safety violations.
In-Fat~\cite{xu2021fat} indexes metadata for heap, stack, and global variables, using embedded metadata, additional registers, and direct-mapped shadow, respectively.
However, In-Fat only implements detection for spatial memory safety bugs, not for temporal security.

\subsection{Memory Safety based on ARM PA}


PARTS~\cite{parts} provides pointer integrity for programs using the ARM PA feature.
It utilizes type information of the target pointers and other auxiliary information as modifiers to sign and authenticate pointers.
Since PARTS uses static type information as the modifier to sign pointers, attackers can bypass it by reusing signed pointers with the same static modifier.

PTAuth~\cite{ptauth} proposes an effective runtime protection scheme for heap-based temporal safety using the ARM PA extension.
PTAuth assigns a unique ID for each object to sign the object's base address.
PTAuth can use the unique ID to do a pointer authentication during every pointer dereference, and report a temporal violation when the authentication fails.
However, this scheme does not protect against spatial violations, e.g., heap overflows, which can overwrite the rest of the heap memory while keeping the ID stored on the heap unchanged.
Besides, PTAuth considers the memory overhead of metadata, so it uses the embedded metadata method to store the ID at the beginning of the object, but this method increases the overhead of querying metadata.
It is necessary to search backward iteratively to find the valid ID when a pointer inside a heap object is being authenticated.
Compared with PTAuth, \sysname's security check is more efficient.

\section{Conclusion}
Existing memory safety sanitizers either provide partial memory safety guarantees or have excessive performance overheads.
Our novel sanitizer \sysname enforces full memory safety
by precisely tracking all necessary memory safety metadata of objects and enforcing complete mediation on all pointer dereferences.
Further, \sysname utilizes the hardware feature ARM PA to seal metadata directly into pointers and places metadata in a global table indexed by the seal.
This mechanism saves the metadata (seal) propagation overhead and enables efficient runtime metadata retrieval and checks.
Experiments demonstrate that \sysname has no false positives and negligible false negatives (i.e., missing checks for sub-object overflows) and provides a stronger security guarantee than state-of-the-art sanitizers,
including HWASan, ASan, SoftBound+CETS, Memcheck, LowFat, and PTAuth, while introducing lower performance overheads.

\clearpage





\bibliographystyle{plain}
\bibliography{main}

\clearpage
\appendix
\renewcommand{\thelstlisting}{\Alph{section}.\arabic{lstlisting}} \renewcommand\thefigure{\thesection.\arabic{figure}}    

\section{Appendix}

\subsection{\bf Comparison between Sanitizers}
\label{sec:appendix-2}

\sysname has advantages over other sanitizers in terms of detecting out-of-bounds access, UAF, and invalid free vulnerabilities.
Next, we will use a specific examples to demonstrate the advantages of \sysname.

\begin{figure}[!h]
\centering
\scriptsize
\begin{lstlisting}[language=C++,frame=shadowbox,label={list:oob},caption={An example of out-of-bounds access.}]
char *buf1 = new char[100];
char *buf2 = new char[1000];
int ptr_offset = 0;
// Leak the address of buf1.
printf("%p\n", (void *)buf1);
// Leak the address of buf2.
printf("%p\n", (void *)buf2);
scanf("%d", &ptr_offset);
buf1[ptr_offset] = 'A';
// 'A' may end up somewhere in buf2.
delete [] buf1;
delete [] buf2;
\end{lstlisting}
\end{figure}

 {\bf Out-of-Bounds Access}\quad
As Listing~\ref{list:oob} shows, if attackers can control the index used to access the \texttt{buf1} array, they can access the memory of \texttt{buf2} and cause a security violation. 
Many sanitizers, including ASan, Memcheck, and LowFat, cannot detect such out-of-bounds access.

Although ASan uses redzones to detect out-of-bounds access, it can only detect a memory operation accessing illegal memory areas, but cannot detect a memory operation jumping over the redzones and accessing another object.
Memcheck only checks whether the accessed address is valid and whether there is an initialized object at the address.
Therefore, it cannot catch out-of-bounds access to a valid and initialized object.
Besides, LowFat utilizes the pointer itself to retrieve the bounds, i.e., LowFat retrieves the bounds using \texttt{(buf1 + ptr\_offset)}. 
It is clear that the bounds retrieved by LowFat is that of \texttt{buf2}, when \texttt{ptr\_offset} is larger than the offset between \texttt{buf2} and \texttt{buf1} (i.e., \texttt{buf2 - buf1}). 
Thus, such out-of-bounds access can also pass the bound check of LowFat.

Since \sysname utilizes base addresses of objects as parts of seals and the seals can propagate along with pointers in a program, \sysname can calculate the accessed offset of the actual referent object for a pointer.
When \texttt{ptr\_offset} is larger than \texttt{buf2 - buf1}, the accessed offset is larger than the size of \texttt{buf1},
thus \sysname can detect such spatial safety violations.

\begin{figure}[!h]
    \centering
\begin{lstlisting}[language=C++,frame=shadowbox,label={list:uaf},caption={An example of use-after-free.}]
char *data1 = (char *)malloc(10 * sizeof(char));
char *data1_copy = data1;
free(data1);
char *a = (char *)malloc(sizeof(char) << 28);  // 256MB
free(a);
char *data2 = (char *)malloc(10 * sizeof(char));
data1_copy[0] = 'A';  // Use after free.
free(data2);
\end{lstlisting}
\end{figure}

{\bf Use After Free}\quad
As Listing~\ref{list:uaf} shows, after deallocating \texttt{data1}, a 256MB object is allocated and freed.
Then, \texttt{data2} is allocated. 
\texttt{data2} has the same size as \texttt{data1} and is located at the same address.
Immediately afterward, the program uses \texttt{data1\_copy}, which is a copy of \texttt{data1}, to store \texttt{'A'} to the memory.
The sanitizers ASan, Memcheck, and LowFat fail to detect such UAF vulnerabilities.
ASan utilizes a quarantine to prevent a just freed object from being immediately reallocated.
However, the quarantine size is limited, which is 256MB in the default configuration.
Thus, ASan cannot detect such UAF vulnerabilities if the quarantine is exhausted and the memory of a previously freed object is reallocated.
Since Memcheck only checks whether the accessed address is valid and whether there is an initialized object at the address, it is also unable to detect such vulnerabilities.
LowFat also fails to detect UAF violations since it focuses only on spatial memory safety.

\sysname clears the metadata of the object pointed by \texttt{data1\_copy} after the deallocation of the object.
Therefore, the seal in \texttt{data1\_copy} has no corresponding metadata in the metadata table and \texttt{data1\_copy} cannot be used to access the referent object (i.e., \texttt{data2}).
Besides, thanks to the unique birthmark for each allocation, 
both the seals of \texttt{data1\_copy} and \texttt{data2} and the metadata of referent objects are different, though the base addresses and the size of them are the same.
Thus, \texttt{data1\_copy} cannot reuse the metadata of \texttt{data2}.
In a word, the birthmark check of \texttt{data1\_copy} will fail, and the attacker cannot use \texttt{data1\_copy} to store \texttt{'A'} to the target memory.

\begin{figure}[!h]
    \centering
\begin{lstlisting}[language=C++,frame=shadowbox,label={list:invalid},caption={An example of invalid free.}]
char *buff;
buff = (char *)malloc(10);
buff++;
free(buff);  // Invalid free.
\end{lstlisting}
\end{figure}

{\bf Invalid Free}\quad
In Listing~\ref{list:invalid}, \texttt{buff} firstly points to a newly allocated object, and then it is increased and does not point to the beginning of that object.
Finally, \texttt{buff} is used to deallocate some object.
Such vulnerabilities are called Invalid Free.

For HWASan, since \texttt{buff} still points to the middle of the object, its tag still matches the tag of the object and can pass the HWASan security checks. 
Similar issues may exist in other tag-based mechanisms including ARM Memory Tagging Extension (MTE).
LowFat cannot detect Invalid Free vulnerabilities as it is designed for detecting spatial safety violations.
\sysname checks whether a pointer used to deallocate some object points is a valid object pointer.
In this case, \texttt{buff} does not point to the beginning of the object, so the check will fail.
In this way, \sysname can detect such vulnerabilities.

\subsection{The result of Magma}
\label{sec:appendix-3}

\begin{table}[htb]
  \footnotesize
    \centering
  \caption{The table includes the number of PoCs provided by Magma (from the AFL++ directory), the number of PoCs  that \sysname detected violations, and the number of PoCs that ASan detected violations.}
    \setlength{\tabcolsep}{0.05mm}{
\begin{tabular}{ccccc}
\hline
open-source libraries     & Test programs                           & PoCs          & \sysname        & ASan          \\ \hline
libpng                    & libpng\_read\_fuzzer                    & 634           & 0             & 0             \\ \hline
\multirow{2}{*}{LibTIFF}  & tiff\_read\_rgba\_fuzzer                & 2197          & 0             & 0             \\
                          & tiffcp                                  & 1519          & 115           & 115           \\ \hline
\multirow{3}{*}{Libxml2}  & libxml2\_xml\_read\_memory\_fuzzer      & 11978         & 0             & 0             \\
                          & libxml2\_xml\_reader\_for\_file\_fuzzer & 0             & 0             & 0             \\
                          & xmllint                                 & 7636          & 0             & 0             \\ \hline
\multirow{3}{*}{Poppler}  & pdf\_fuzzer                             & 1803          & 190           & 190           \\
                          & pdfimages                               & 3455          & 659           & 659           \\
                          & pdftoppm                                & 2085          & 480           & 480           \\ \hline
\multirow{12}{*}{OpenSSL} & asn1                                    & 55            & 14            & 14            \\
                          & asn1parse                               & 0             & 0             & 0             \\
                          & bignum                                  & 0             & 0             & 0             \\
                          & bndiv                                   & 0             & 0             & 0             \\
                          & client                                  & 194           & 140           & 140           \\
                          & cmp                                     & 0             & 0             & 0             \\
                          & cms                                     & 0             & 0             & 0             \\
                          & conf                                    & 0             & 0             & 0             \\
                          & crl                                     & 0             & 0             & 0             \\
                          & ct                                      & 0             & 0             & 0             \\
                          & server                                  & 62            & 30            & 30            \\
                          & x509                                    & 344           & 10            & 10            \\ \hline
SQLite                    & sqlite3\_fuzz                           & 1777          & 0             & 0             \\ \hline
\multirow{5}{*}{PHP}      & exif                                    & \textbf{1443} & \textbf{1128} & \textbf{1083} \\
                          & json                                    & 0             & 0             & 0             \\
                          & mbstring                                & 0             & 0             & 0             \\
                          & parser                                  & 0             & 0             & 0             \\
                          & unserialize                             & 0             & 0             & 0             \\ \hline
Lua                       & lua                                     & 0             & 0             & 0             \\ \hline
libsndfile                & sndfile\_fuzzer                         & 0             & 0             & 0             \\ \hline
\end{tabular}
}
\label{tab:detailed_magma}
\end{table}

Magma includes 9 common open-source libraries and have some test programs for each libraries. In Table~\ref{tab:detailed_magma}, we detail the detection of PoCs for each of the test programs.

As can be seen from Table~\ref{tab:detailed_magma}, \sysname and Asan detect the same number of PoCs for most programs.
For PHP, \sysname found 45 more PoCs than ASan, all of which were PoCs from CVE-2018-14883.
CVE-2018-14883 is a heap-based buffer over-read caused by an Integer Overflow. 
The 45 more PoCs found by \sysname than ASan all crossed the RedZone of ASan to access another valid memory object when read memory. 
Therefore, ASan cannot detect the violation in these PoCs.
Specifically, these 45 PoCs had a total of three cases of out-of-bounds access, i.e., memory violation was acted by three different instructions. Two situations of the out-of-bounds access were in the {\tt php\_ifd\_get32u} function, which used the vulnerable pointers for array accesses. The other situation is a call instruction of the {\tt memcpy} function, which uses the vulnerable pointer to copy 6 bytes of the targeted object.

\subsection{The omited test cases in Juliet}
\label{sec:appendix-juliet}
We analyzed the test cases and found that some test cases can cause fake false negatives, which fall into the following categories:

First, some type-confusion test cases do not trigger buffer overflows, such as {\tt Heap\_Based\_Buffer\_Overflow\_\_sizeof\_double*}.
Specifically, these test cases allocate memory to an object with the pointer size rather than the object size, while the object's size should be 64 bits (such as a {\tt double} type).
And then, it accesses the target object using the object size rather than the allocated size.
On a 64-bit system like ARM64, the pointer size and the object size are both 64 bits, and there is no buffer overflow.

Second, some test cases rely on a random number to trigger the vulnerability and may not trigger out-of-bounds access during testing.

Moreover, there are 18 test cases in the {\tt CWE476} class called {\tt NULL\_Pointer\_Dereference\_\_null\_check\_after\_deref*}.
The only difference between the \texttt{BAD} and \texttt{GOOD} programs of these test cases is that each \texttt{BAD} program has an extra null-pointer check after the target pointer has been dereferenced. 
There are annotations of the source code of these test cases, which said: "This NULL check is unnecessary."
We believe that these 18 test cases are used to evaluate the accuracy of static analysis tools and cannot trigger the vulnerability in practice. 
Therefore, we omit the results of these 18 \texttt{BAD} programs when calculating false-negative rates.

Since these cases do not have out-of-bound access on the testing platform, we do not count them as false negatives for memory safety sanitizers.

\subsection{Artifact Description}
Apple claimed that M1 includes 4 high-performance cores and 4 high-efficiency cores. But we found that there are 3 types of cores in M1 mini by testing the time overhead of the {\tt ADD} instruction.
Specifically, on cores 0, an {\tt ADD} instruction requires 0.714 ns; on cores 1-3, an {\tt ADD} instruction requires 0.485 ns; on cores 4-7, an {\tt ADD} instruction requires 0.335 ns.
To be fair, we need to use the {\tt numactl} command to ensure that the target programs are only running on cores 4-7.

In summary, the experimental environment used for our prototype system was as follows:
\begin{enumerate}
\item System Architecture: AArch64 ARMv8.3+
\item Computer System: Linux OS
\item Device used: MAC M1 mini
\item CPU id used for performance testing: 4-7
\end{enumerate}

\end{document}